\documentclass[aps,prb,superscriptaddress,twocolumn]{revtex4}

\usepackage{bm}
\usepackage{graphicx}
\usepackage{amsmath}
\usepackage{subfigure}

\usepackage{amssymb,amsfonts,amsmath}

\DeclareMathOperator*{\Tr}{Tr}

\DeclareRobustCommand\openzero{\leavevmode\hbox{0\kern-.55em0}}
\newcommand{\tildeS}{{\widetilde S}}

\newcommand{\ket}[1]{|{#1}\rangle}
\newcommand{\Ket}[1]{|{#1}\rangle}
\newcommand{\bra}[1]{\langle{#1}|}
\newcommand{\scalar}[2]{\langle{#1}|{#2}\rangle}
\newcommand{\exval}[1]{\langle{#1}\rangle}

\RequirePackage{color}

\begin{document}



\title{Parametric representation of open quantum systems and crossover 
from quantum to classical environment}




\author{D. Calvani}
\affiliation{Dipartimento di Fisica, Universit\`a di Firenze,
             Via G. Sansone 1, I-50019 Sesto Fiorentino (FI), Italy}
\affiliation{INFN Sezione di Firenze, via G.Sansone 1,
             I-50019 Sesto Fiorentino (FI), Italy}
\author{A. Cuccoli}
\affiliation{Dipartimento di Fisica, Universit\`a di Firenze,
             Via G. Sansone 1, I-50019 Sesto Fiorentino (FI), Italy}
\affiliation{INFN Sezione di Firenze, via G.Sansone 1,
             I-50019 Sesto Fiorentino (FI), Italy}
\author{N.I. Gidopoulos}
\affiliation{ISIS, STFC, Rutherford Appleton Laboratory, Didcot, OX11 0QX, 
United Kingdom }
\author{P. Verrucchi}
\affiliation{Istituto dei Sistemi Complessi, Consiglio Nazionale delle Ricerche,
             via G. Sansone 1, I-50019 Sesto Fiorentino (FI), Italy}
\affiliation{Dipartimento di Fisica, Universit\`a di Firenze,
             Via G. Sansone 1, I-50019 Sesto Fiorentino (FI), Italy}
\affiliation{INFN Sezione di Firenze, via G.Sansone 1,
             I-50019 Sesto Fiorentino (FI), Italy}



\begin{abstract} The behaviour of most physical systems is affected by 
their natural surroundings. A quantum system with an environment is 
referred to as "open", and its study varies according to the classical 
or quantum description adopted for the environment. We propose an 
approach to open quantum systems that allows us to follow the crossover 
from quantum to classical environments; to achieve this, we devise an 
exact parametric representation of the principal system, based on
generalized coherent states for the environment. The method is applied 
to the $s{=}\frac{1}{2}$ Heisenberg-star with frustration, where the 
quantum character of the environment varies with the couplings entering 
the Hamiltonian $H$. We find that when the star is in an eigenstate of 
$H$, 
the central spin behaves as if it were in an effective magnetic field,
pointing in the direction set by the environmental coherent-state angle
variables $(\theta,\varphi)$, and broadened 
according to their quantum probability distribution. Such distribution 
is independent on $\varphi$ while, as a function of $\theta$, is seen to 
get narrower as the quantum character of the environment is reduced, 
collapsing into a Dirac-$\delta$ function in the classical limit. In such 
limit, as $\varphi$ is left undetermined, 
the Von Neumann entropy of the central spin remains finite. It is equal to
the entanglement of the original fully-quantum model, which further
establishes a relation between this latter quantity and the Berry-phase
characterizing the dynamics of the central spin in the effective magnetic
field.
\end{abstract}
\maketitle

\section{Introduction}
Quantum systems with an environment are usually referred to as {\em open 
quantum systems} (OQS). Despite having been extensively studied since 
the very birth of quantum  mechanics, the relevance acquired by 
certain features (such as coherence and entanglement) in the last decades, have 
boosted further the interest towards their behaviour, both in the 
dissipative and in the non-dissipative case
\cite{LeggetEtal87,Weiss99,BreuerP02, RivasH11}.
The description of an OQS in terms of the corresponding reduced 
density matrix, defined as the partial trace of the density matrix
of the global system over the Hilbert space of the environment, 
is axiomatically exact and fully retains the 
quantum character of the environment, making it the preferred 
approach, particularly in the realm of quantum information theory 
and computation \cite{BennetS98,NielsenC00,KayeLM07,Mermin07}. However, there 
exists 
another description of OQS, where the principal system is in a pure state, but 
under the effect of a local Hamiltonian depending on ``external" 
parameters, whose presence testifies the existence of 
a surrounding environment. At the heart of this approach stands the 
approximation that the environment be classical, so that the operators 
acting on its Hilbert space are
replaced by c-number parameters; in this way, the 
interaction Hamiltonian is reduced to an effectively local one for the 
sole principal system. The prototypical example 
of this theoretical scheme is that of a spin in an external magnetic field, 
but many other examples can be easily spotted. 

In the two approaches sketched above, the environment is either quantum 
or classical. Aim of this work is to devise a method for studying OQS in 
the non-dissipative case, capable of interpolating between the quantum 
description of the environment and its classical limits, without 
affecting the quantum nature of the principal system.
Inspired by the Born-Oppenheimer approach\cite{BornO27}, 
where electrons are described in terms of pure states parametrically 
dependent on the nuclear positions, we construct a representation of OQS 
in terms of pure states that depend on continuous parameters 
characterising the environment. Such representation is formally exact, 
in the same way that the parametric separation between electronic and 
nuclear subsystems can be made exact\cite{Hunter75,GidopoulosG05}, with 
the role of principal system and environment essentially 
interchangeable.
Once the procedure for constructing the above parametric representation, 
which is based on the use of generalized coherent states for the 
environment, is devised, we implement it in the case of a specific spin 
model, namely the spin-$\frac{1}{2}$ star with 
frustration\cite{RichterV94,DengF08}: this system belongs to the large 
family of the so-called "central-spin" models, which have been 
extensively studied 
\cite{ProkofievS00,BreuerBP04,HuttonB04,Al-HassaniehEtal06,PalumboNM06,%
LibertiEtal06,HamdouniP07,KroviEtal07, 
HansonEtal08,FerraroEtal08,BortzEtal10}, since they describe 
magnetic interactions that play a relevant role in the physics of 
candidate future nanodevices 
\cite{LossDV98,Kane98,ImamogluEtal99,SchliemannKL03,CoishL04,%
RossiniEtal07,CywinskiWDS09,HaikkaetAl12}.

The parametric representation, besides providing an insight 
into the physical behaviour of the above
spin model, prove suitable for dealing with peculiar geometrical effects 
such as the Berry's phase\cite{Simon83,Berry84}, 
and allows us to confront a very relevant 
topic pertaining OQS, namely the subtle connection between geometric 
aspects of quantum  mechanics and entanglement\cite{Levay04,Chruscinski06}, 
as discussed in the final part of the paper.

\section{The parametric representation with coherent states} 
\label{s.parametricrep}
Let us consider 
an isolated system in a pure, normalized, state $\ket{\Psi}$: the system is 
made of a principal system $A$ and its environment $B$, with 
separable Hilbert spaces 
$\mathcal{H}_A$ and $\mathcal{H}_B$, respectively. 
Given two orthonormal bases $\{\ket{\alpha}\}\subset\mathcal{H}_A$ and 
$\{\ket{\beta}\}\subset\mathcal{H}_B$, it is 
$\ket{\Psi}=\sum_{\alpha\beta}c_{\alpha\beta}\ket{\alpha}\ket{\beta}$.
Aimed at the interpolation scheme mentioned in the 
Introduction, 
we understand that describing the environment in terms of continuous parameters,
rather than with a set of discrete ones, can be rewarding.
The idea of introducing proper ``environmental coherent states" 
naturally follows, 
and we accomplish it by resorting to the group-theoretical construction 
proposed by Gilmore (see Ref.\cite{ZhangFG90} and references therein).
Let the total environmental Hamiltonian be
$H_B+H_{AB}$, where the first, local, term contains 
operators acting on $\mathcal{H}_B$, while the second, interaction, 
term contains operators acting on $\mathcal{H}_A\otimes \mathcal{H}_B$. 
We assume $H_{AB}$ is a linear combination of tensor products of 
operators acting on $\mathcal{H}_A$ and on 
$\mathcal{H}_B$, as is the case in most physical situations.
Let $G$ be the dynamical group in terms of whose generators we 
can write the environmental Hamiltonian; 
the Hilbert space $\mathcal{H}_B$ is given by the specific 
physical set up and is associated with a unitary irreducible 
representation of $G$. The choice of a reference state 
$\ket{\beta_0}\in \mathcal{H}_B$ fixes: {\it i)} 
the maximum stability subgroup $F\subset G$, i.e. the set of those 
operators $f$ such that $f\ket{\beta_0}=e^{i\lambda_f}\ket{\beta_0}$; {\it ii)} 
the quotient $G/F$, such that any $g\in G$ can be 
locally decomposed as $g=\Omega f$, with $f\in F$ and $\Omega\in G/F$.
Generalized {\it coherent states} $\ket{\Omega}$ are eventually defined as 
$\ket{\Omega}=\Omega\ket{\beta_0}$, with $\Omega\in G/F$;
they can be normalized, and form an overcomplete set in $\mathcal{H}_B$, \emph{i.e.} 
\begin{equation}
 \int\text{d}\mu(\Omega)\Ket{\Omega}\bra{\Omega}={\bm{1}}_{\mathcal{H}_B}\,\,,
\label{e.continuous-identiy}
\end{equation}
with $\text{d}\mu(\Omega)$ a proper group-invariant measure in $G/F$, 
and $\exval{\Omega|\Omega'}\neq\delta(\Omega-\Omega')$.
Using the above resolution of the identity in $\mathcal{H}_B$, the state 
of the total system can be written as
\begin{equation}
\Ket{\Psi}=\int\text{d}\mu(\Omega)
\chi^\Psi(\Omega)\Ket{\Omega}\Ket{\phi_A^\Psi(\Omega)}\,\,,
\label{e.Psipara}
\end{equation}
with
\begin{eqnarray}
&\!&\ket{\phi^\Psi_A(\Omega)}\equiv\frac{1}{\chi^\Psi(\Omega)}
\sum_\alpha f_{\alpha}(\Omega)\ket{\alpha}
\,\,,\label{e.phicoherent}\\
&\!&f_\alpha(\Omega)\equiv\sum_{\beta}\bra{\Omega}\beta\rangle c_{\alpha\beta}
\,\,,\\
&\!&\chi^\Psi(\Omega)\equiv e^{i\Lambda(\Omega)} 
\sqrt{\sum_{\alpha}|f_\alpha(\Omega)|^2}
\,\,,\label{e.chicoherent}
\end{eqnarray}
where $e^{i\Lambda(\Omega)}$ is a gauge freedom.
Each ket $\ket{\phi^\Psi_A(\Omega)}$ is a 
normalized element of $\mathcal{H}_A$ and therefore represents a physical state for 
the principal system.

By parametric representation of $A$ we will hereafter mean its description 
in terms of the pure states 
$\{\ket{\phi^\Psi_A(\Omega)}\}$: 
the dependence of  $\ket{\phi^\Psi_A(\Omega)}$ on the parameter $\Omega$ 
is the fingerprint that an environment exists.
Actually, it can be easily shown that $\ket{\phi_A^\Psi}$ 
depends on the parameter $\Omega$ if and only if the 
global state $\Ket{\Psi}$ is entangled:
it is the entangled structure of $\Ket{\Psi}$ that causes the
dependence on $\Omega$ to be conveyed from the environment to the 
principal system, a fact that will play a crucial role in relating 
entanglement and geometrical phases, as discussed in the last Section.

The  normalization of 
$\ket{\Psi}$ implies $\int\text{d}\mu(\Omega)|\chi^\Psi(\Omega)|^2=1$, 
allowing $|\chi^\Psi(\Omega)|^2$ to be interpreted as a probability 
distribution 
on $G/F$, which turns out to be the phase space of the 
environment  under rather general assumptions\cite{ZhangFG90}; 
indeed, by a simple calculation it is immediate to 
show that $|\chi^\Psi(\Omega)|^2$ is just the Husimi $Q$-function of the 
environmental reduced density matrix \cite{ZhangFG90, Lieb73}, 
$\exval{\Omega| (\Tr_A\ket{\Psi}\bra{\Psi})|\Omega}=|\chi^\Psi(\Omega)|^2$. 

The resulting picture for the principal system is that of a continuous 
collection of pure, parametrized, states whose occurrence is ruled by a 
probability distribution 
over the phase space of the environment. The relation
$\text{Tr}_B [\,\cdot\,]=\int \text{d}\mu(\Omega) \exval{\Omega| 
\,\cdot\,|\Omega}$ holds \cite{Lieb73}, yielding, for the reduced 
density matrix of the principal system, $\rho_A=\int 
\text{d}\mu(\Omega)|\chi(\Omega)|^2
\Ket{\phi^\Psi_A(\Omega)}\bra{\phi^\Psi_A(\Omega)}$, and hence 
\begin{equation}
\exval{O}\equiv\text{Tr}_A (\rho_A O)= 
\int\text{d}\mu(\Omega)
|\chi(\Omega)|^2\exval{\phi^\Psi_A(\Omega)|O|\phi^\Psi_A(\Omega)}\,\,,
\label{e.expecval}
\end{equation} 
for any principal-system observable $O$.
The above construction provides an exact description for both 
the principal system and its environment, and further allows us
to treat them in a very different formal scheme: in fact, and at 
variance 
with other related works\cite{Al-HassaniehEtal06,BoixoVO07,ShalashilinC00}, 
coherent states are here exclusively adopted for the environment.
This makes the approximations which naturally arise in the 
coherent state formalism available for describing the environment,
but it also prevents the principal system from being affected by 
those same approximations.

Before ending this section, we notice that any resolution of the 
identity in the Hilbert space of the environment defines a
parametric representation for the principal system. In fact, the 
construction 
of the parametric representation described above can 
be both generalized, referring to approaches to quantum 
mechanics on phase space
\cite{Schroeck96} that go beyond the theory of coherent states, 
and made more specific, whenever continuous 
variables different from those related to coherent states
emerge in dealing with specific types of environment 
\cite{AdessoI07,VasileEtal10}.

\section{The Spin-$\frac{1}{2}$ star with frustration.}
We apply the above formalism to the specific physical situation 
where a spin $1/2$ (${\bm\sigma}/2$, hereafter 
called \emph{qubit}) interacts with an even set of $N$ 
spins $1/2$ (${\bm s}_i$, hereafter called \emph{environmental 
spins}) \emph{via} an isotropic antiferromagnetic 
coupling \cite{RichterV94,DengF08}. 
The environmental 
spins interact among themselves and the total Hamiltonian 
is that of the so-called ``spin-$\frac{1}{2}$ star with frustration", 
\begin{eqnarray}
&~&H=H_B+H_{AB}\,\,,\label{e.Hspinstar}\\
&~&H_B=\frac{k}{N}\sum_i^N{\bm s}_i\cdot{\bm s}_{i+1}
\,\,\,;\,\,\,\,\,\,k>0
\label{e.HB}\\
&~&H_{AB}=\frac{g}{N}\frac{\bm\sigma}{2}\cdot\sum_i{\bm s}_i
\,\,\,\,;\,\,\,\,\,\,g>0
\label{e.HAB}
\end{eqnarray}
where $i$ runs over the sites of the external ring.
As mentioned in the introduction, the spin-$\frac{1}{2}$ star with 
frustration belongs to a class of ``central-spin"-like models that have 
been extensively studied in the last decade.
Without entering a detailed case study, which goes beyond the scope 
of this paper, we underline that this model posseses the most welcome 
property of 
allowing the quantum character of the 
environment, as measured by the total spin of the ring, to be varied 
acting on the ratio $k/g$ (usually referred to as the {\it frustration 
ratio}), as discussed below. Moreover the model is exactly
solvable and analytical expressions of its eigenstates and eigenvalues 
are available \cite{RichterV94}.

Let us briefly review the main known facts about the spin-$\frac{1}{2}$
star with frustration, Hamiltonian 
\eqref{e.Hspinstar}-\eqref{e.HAB}: the integrals of 
motion are the total Hamiltonian $H$, the local environmental Hamiltonian 
$H_B$, the square ${\bm J}^2$ of the total angular momentum ${\bm 
J}\equiv{\bm\sigma}/2+\sum_i {\bm s}_i$, its component along the 
quantization axis, $J^z$, and the square ${\bm S}^2$ of the total spin 
of the ring $\bm S=\sum_i {\bm s}_i$,
with the respective eigenvalues $E,\,kE_B,\,J(J+1),\,M,$ and $S(S+1)$;
it is also useful to define $m$ as the eigenvalues of the component 
$S^z$ of the total environmental spin ${\bm S}$.
The relations $J=S\pm\frac{1}{2}$ and $M=m\pm\frac{1}{2}$ hold.
The total spin $S$ ranges from $0$ to $N/2$. 
Apart from the state with $S=0$, which has energy $kE_B$, for each 
assigned value $S>0$ the energy spectrum consists of the two multiplets
\begin{eqnarray}
E_+&=&kE_B+\frac{g}{2N}S\,\,,\label{e.E+}\\
E_-&=&kE_B-\frac{g}{2N}(S+1)\,\,;
\label{e.E-}
\end{eqnarray}
the lowest eigenvalue of $H_B$ obeys the Lieb-Mattis ordering 
\cite{LiebM62}, $E_B(S)<E_B(S+1)$, and the competition between the 
two terms in Eq.~\eqref{e.E-}, embodying the antiferromagnetic frustration 
of the model, makes the ground-state (GS) of the star belong to the 
$E_-$ multiplet with a value of $S$ that varies with the frustration ratio 
$k/g$. In fact, for $0\le k/g\le 1/4\equiv\alpha_0$, 
the GS 
has $S=N/2$ while, when $k/g$ increases, there exists a sequence of 
critical values $\alpha_n$, $n=1,...,(N/2-1)$, such that $S=N/2-n$ for 
$\alpha_{n-1}<k/g\le\alpha_{n}$, and $S=0$ for 
$k/g>\alpha_{N/2-1}>>1$. Notice that the 
critical values $\alpha_n$ depend on $N$, with the exception of 
$\alpha_0$ which equals $1/4$ for all $N$. 
The basic structure of the eigenstates belonging to a subspace with fixed 
$S$ is
\begin{equation}
\ket{\Psi^\pm_{M}}=a^\pm_M\ket{\uparrow}\ket{m_-}+
b^\pm_M\ket{\downarrow}\ket{m_+}~,
\label{e.eigenstates}
\end{equation}
where $\ket{\uparrow,\downarrow}$ are the eigenstates of $\sigma^z$,
$\ket{m_\pm}$ are the eigenstates of $S^z$ with eigenvalues 
$m=M\pm 1/2$, and the apex $\pm$ refers to the state having 
energy $E_\pm$.  
Introducing $\tildeS\equiv S+1/2$, the 
coefficients are
\begin{equation}
a^\pm_M=\pm\frac{1}{\sqrt{2}}\sqrt{1\pm\frac{M}{\tildeS}}\,\,,\,\,
b^\pm_M=\frac{1}{\sqrt{2}}\sqrt{1\mp\frac{M}{\tildeS}}\,\,,
\label{e.ab}
\end{equation}
yielding $a^\pm=\pm b^\mp$.
The entanglement between the qubit and its environment for the states 
\eqref{e.eigenstates}, as measured by the Von Neumann entropy, is 
easily found to be the same for both multiplets and to depend just on 
the ratio $M/\tildeS\equiv\cos\vartheta_M$, according to 
\begin{equation}
{\cal{E}}_{\sigma S}=
-h\left[\frac{1}{2}\left(1-\cos\vartheta_M\right)\right]\,\,,
\label{e.entanglement}
\end{equation}
where $h[x]$ ($0\le x\le 1$) is the binary entropy $x\log x + (1-x)\log 
(1-x)$. 

We construct the parametric representation of the states 
\eqref{e.eigenstates}, each having $S$ fixed, 
by identifying ${\bm S}=\sum_i{\bm s}_i$ with the environment and 
$\bm\sigma$ with the principal system;
the dynamical group $G$ to be employed in the generalized coherent 
state construction is therefore $SU(2)$, with $\mathcal{H}_B$ its 
spin$-S$ irreducible representation. We choose the reference state 
$\Ket{\boldsymbol{\beta}_0}$ to be the maximal weight of the spin 
$S$-representation, \emph{i.e.} $\Ket{\boldsymbol{\beta}_0}= \Ket{m=S}\in \mathcal{H}_B$, so 
that $U(1)$ is the maximum stability subgroup $F$, eventually ending up in 
the \emph{Bloch coherent states} 
$\Ket{\Omega}$, whose expansion over the $S_z$ basis 
$\{\Ket{m}\}\in \mathcal{H}_B$ is (using the conventions of \cite{Lieb73})
\begin{equation}\label{e.Omega} 
\Ket{\Omega}\equiv
\displaystyle{e^{\zeta S^--\zeta^*S^+}}\Ket{m=S}=\sum_{m=-S}^S
g_{m}(\theta)e^{\textup{i}(S-m)\varphi}\Ket{m} \,\,,
\end{equation} 
where $S^{\pm}=S_x\pm iS_y$, $\zeta={\theta\over2}e^{i\varphi}$, 
\begin{equation}\label{e.gtheta} 
g_{m}(\theta)\equiv\sqrt{\binom{2S}{S+m}}\cos^{S+m}
\left(\frac{\theta}{2}\right)\sin^{S-m}\left(\frac{\theta}{2}\right)\,\,, 
\end{equation}  
and $\Omega\equiv(\theta,\varphi)$ is a point of the 
$S^2\simeq SU(2)/U(1)$ sphere.
The resolution of the identity reads
${\bm 1}_{\mathcal{H}_B}=\frac{\tildeS}{2\pi}\int \textup{d}\Omega 
\Ket{\Omega}\bra{\Omega}$, where
$\textup{d}\Omega\equiv\sin\theta\text{d}\theta\text{d}\varphi$ is the 
euclidean measure on the $S^2$ sphere.
From the definitions \eqref{e.Psipara}-\eqref{e.chicoherent} and the expansion 
\eqref{e.Omega}, we 
can explicitely write the states \eqref{e.eigenstates} as
\begin{equation}\label{e.eigenpar}
\Ket{\Psi^\pm_M}=\frac{\tildeS}{2\pi}\int 
\text{d}\Omega\,\chi^{\pm}_M(\Omega)\Ket{\Omega}\Ket{\phi^{\pm}_M(\Omega)}\,\,,
\end{equation}
where
\begin{equation}
\chi^\pm_M(\Omega)=e^{\textup{i}(\tildeS-M)\varphi}
\sqrt{[a^\pm_M g_{M-\frac{1}{2}}(\theta)]^2+
[b^\pm_Mg_{M+\frac{1}{2}}(\theta)]^2}\,\,,
\label{e.chiS}
\end{equation}
and the qubit parametrized states are
\begin{equation}
\Ket{\phi^\pm_M(\Omega)}=
\pm\cos\left(\frac{\Theta^\pm_M(\theta)}{2}\right)\Ket{\uparrow}+
\sin\left(\frac{\Theta^\pm_M(\theta)}{2}\right)e^{\textup{i}\varphi}
\Ket{\downarrow}\,\,,
\label{e.qubitpar}
\end{equation}
where $\Theta^\pm_M(\theta)$ are the solution with respecto to 
$\Theta$ of the equations
\begin{equation}
\tan\frac{\Theta}{2}=
\left(\tan\frac{\vartheta_M}{2}\right)^{\pm 1}
\tan\frac{\vartheta_M}{2}\cot\frac{\theta}{2}\,\,.
\label{e.Theta}
\end{equation}
Notice that although $\ket{\Omega}$ depends on $\varphi$ according to
Eq.\eqref{e.Omega}, the distribution $|\chi^\pm_M(\Omega)|^2$ does not.
This follows from the structure of the eigenstates 
\eqref{e.eigenstates},
implying that only one term enters the sum in Eq.\eqref{e.chicoherent} 
and no interference effect consequently emerges in $|\chi^\pm_M(\Omega)|^2$.
Such feature is specifically due to the symmetry of the Heisenberg 
interaction and plays an essential role when the classical limit of the 
ring is taken, as thoroughly discussed in the next Section.

The parametric representation of the qubit is given by 
the kets \eqref{e.qubitpar}, each corresponding to the physical state the qubit is in, provided the total system is in $\Ket{\Psi^\pm_M}$ 
and the environment in the coherent state $\ket{\Omega}$.
As $|\chi^\pm_M(\Omega)|$ does not depend on $\varphi$,
the $\Omega$-occurence is ruled  by the $\theta$-normalized probability 
distribution $\tildeS\sin(\theta)|\chi^\pm_M(\Omega)|^2\equiv 
p_M^\pm(\theta)$, which can be shown to be the properly normalized 
$Q$-representation of the environmental density matrix.
\begin{figure}[h]
\centering 
{\includegraphics[width=0.45\textwidth]{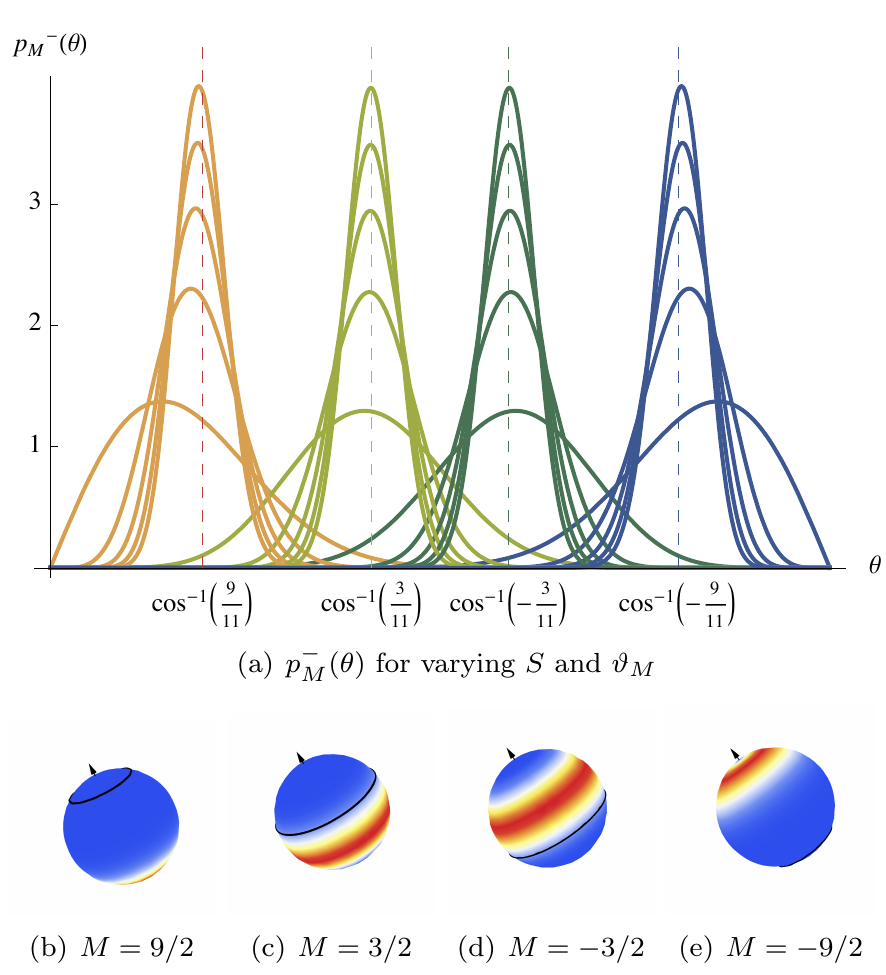}}\\ 
\caption{Upper panel, (a): $p^-_M(\theta)$ for 
$\frac{M}{\tildeS}=\frac{9}{11},\frac{3}{11},-\frac{3}{11},-\frac{9}{11}$
(from left to right) and $S$=5,16,27,38,49 (from below); the vertical 
dashed lines mark the corresponding values of $\vartheta_M$ (see text), 
each identified by a given color. Lower panel, (b-e): qubit-states
distributions, $\pi^-_M(\pi-\theta,\varphi)$, on the Bloch sphere for $S=5$ and 
$\vartheta_M$ as in the upper panel. A black line of 
latitude marks $\vartheta_M$ on each sphere, and the corresponding value of 
$M$ is reported below.}
\label{f.fig1}
\end{figure}
\noindent In the upper panel of Fig.\ref{f.fig1} we show
$p^-_M(\theta)$ for different values of $M/\tildeS$;
the values of $S$ are chosen so as to consistently 
correspond to half-integer $M$.
Two effects clearly testify that we are dealing with a quantum 
environment: the finite width of the distribution and the position shift 
of its maximum with respect to $\vartheta_M$, such shift being the signature 
that the qubit exists. When the quantum character 
of the environment is reduced, \emph{i.e.} $S$ is increased with $M/\tildeS$ 
fixed, both quantities lessen. 
\noindent The probability distribution of the qubit states \eqref{e.qubitpar}
on the Bloch sphere, parametrized by $\Theta$ and $\varphi$, 
is the integrand of Eq.\eqref{e.expecval} with 
$O=\ket{\phi^\pm_M}\bra{\phi^\pm_M}$, and reads 
$\pi^\pm_M(\Theta,\varphi)=p^\pm_M(\theta)$, 
with $\Theta$ and $\theta$ connected by Eq.\eqref{e.Theta}. 
When the total system is in its GS multiplet, it is easily seen that 
$\Theta=\pi-\theta$ for all values of $M$. 
The lower panel of Fig.\ref{f.fig1} shows
$\pi^-_M(\Theta,\varphi)=p^-_M(\pi-\theta)$ 
for $M/{\widetilde S}$ as in the upper panel,
and $S=5$ (each sphere is below the 
corresponding distribution); the angle $\vartheta_M$ is identified by the 
black line of latitude.
\begin{figure}[h]
\includegraphics[width=0.45\textwidth]{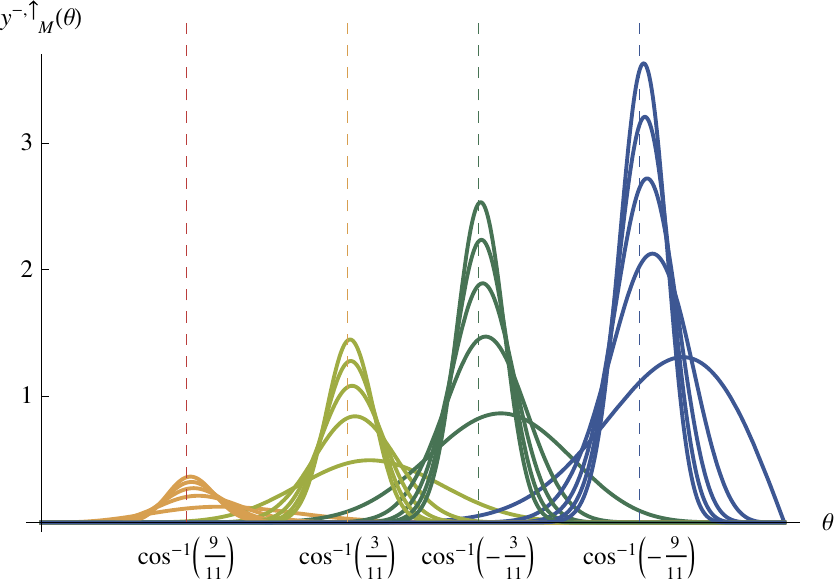}
\caption{Conditional probability distributions for the qubit to be
in the $\ket{\uparrow}$ state when the total system is in its ground 
state. Values of $S$ and $M/\tildeS$, as well as dashed lines meaning, 
as in Fig.~1.}
\label{f.fig2}
\end{figure}
\noindent The conditional probability distribution for the qubit to be, 
say, in the state $\ket{\uparrow}$, reads
\begin{equation}
y^{\pm,\uparrow}_M(\theta)=p^\pm_M(\theta)\cos^2\frac{\Theta_M^\pm(\theta)}{2}\,;
\label{e.condprobup}
\end{equation} 
In the case of the GS multiplet, despite 
$\ket{\phi^-_M(\Omega)}$ being independent on $M$, 
due to $\Theta_M^-=\pi-\theta$ for all $M$, 
the conditional probability distribution \eqref{e.condprobup} inherits 
the dependence on $M$ from the environment,
as cleary seen in Fig.\ref{f.fig2}. 
Notice, that such dependence accounts for the antiferromagnetic character of 
the interaction term \eqref{e.HAB} by representing the 
counteralignment of the qubit with respect to the total environmental spin.

Let us now further comment on the parametrized qubit 
states \eqref{e.qubitpar}: 
it is easily shown that one can always define magnetic fields such that 
the kets $\ket{\phi^\pm_M(\Omega)}$ are the 
ground (``-") and excited (``+") states of the corresponding Zeeman 
terms. In fact, defining
${\bm h}^\pm_M(\Omega)\equiv
\varepsilon^\pm{\bm n}(\Theta^\pm_M,\varphi)$, with 
${\bm n}(\Theta^\pm_M,\varphi)\equiv
(\sin\Theta_M^\pm\cos\varphi,\sin\Theta_M^\pm\sin\varphi,\cos\Theta^\pm_M)$
and $\Theta^\pm_M$ from Eq.\eqref{e.Theta}, it 
is easily verified that 
${\bm h}^\pm_M\cdot\frac{\bm\sigma}{2}\ket{\phi^\pm_M(\Omega)}=
\frac{\varepsilon^\pm}{2}\ket{\phi^\pm_M(\Omega)}$, 
for all $M$. By further requiring
\begin{equation}
\int\text{d}\mu(\Omega)|\chi(\Omega)|^2
\exval{\phi^\pm_M(\Omega)|
{\bm h}^\pm_M\cdot\frac{\bm \sigma}{2}|\phi^\pm_M(\Omega)}=
\bra{\Psi^\pm_M}H_{AB}\ket{\Psi^\pm_M}\,,
\end{equation}
one can also fix the 
field intensity, $\varepsilon^\pm=\pm\frac{g}{N}(\widetilde{S}\mp 1)$.
The above result means that when the star is 
in one of its eigenstates $\ket{\Psi^\pm_M}$
the central spin can be described by a continuous 
collection of pure states which are the eigenstates of 
${\bm h}^\pm_M(\Omega)\cdot\frac{\bm \sigma}{2}$, 
with the direction of the fields distributed according to 
the environmental quantum probability $|\chi^\pm_M(\Omega)|^2$.
This also provides a key to the reading of Fig.\ref{f.fig1}, whose
lower panel visualizes the qubit response to the application of a field
with direction distributed as shown in the upper panel. 
Picturing the qubit in terms of the above "effective" magnetic 
fields, one is naturally led to the last part of our work, namely the crossover 
from a quantum to a classical environment.

\section{The large-$S$ limit.}
From a direct calculation one can see that
\begin{equation}
p_M^\pm(\theta)\xrightarrow{S\to\infty} \delta(\theta-\vartheta_M)\,\,,
\label{e.rhocl}
\end{equation}
meaning that the variable describing the polar angle of the 
environmental coherent states, $\theta$, gets frozen to the value 
$\vartheta_M$, which is fixed by the state of the global system, 
$\ket{\Psi^\pm_M}$.
The qubit parametrized states consequently collapse into the kets 
\eqref{e.qubitpar} with $\Theta^+_M=\vartheta_{M}$ and 
$\Theta^-_M=\pi-\vartheta_{M}$,
hereafter indicated by $\ket{\phi^\pm_M(\varphi)}$.
Notice that the dependence on $\theta$ is fully removed by the 
above classical limit, but all the azimuthal angles 
$\varphi\in[0,2\pi)$ are still allowed.

The kets $\ket{\phi^\pm_M(\varphi)}$ are the stationary states of a 
qubit in a field $\frac{g}{2}{\bm n}^\pm_M(\varphi)$, with 
${\bm n}^\pm_M(\varphi)\equiv {\bm n}(\Theta^\pm_M,\varphi)$ and
$\Theta^\pm_M$ as above (notice that $S\to\infty$ implies $S/N\to 1/2$, 
so that $\varepsilon^\pm$ converge to $g/2$).
From this perspective, the model after 
Eq.\eqref{e.rhocl} is that of a qubit in a pure state, 
which is one of the eigenstates of a local parametric Hamiltonian
$\frac{g}{2}{\bm n}^\pm_M(\varphi)\cdot\frac{\bm \sigma}{2}$, where the only
parametric dependence left is that on the azimuthal angle of the field.
Put it this way, with the qubit in a pure state, one seems to loose the 
connection with the original composite quantum system and, in 
particular, with the fact that the parametric dependence is indeed a 
consequence of the entangled structure of its
global state $\ket{\Psi^\pm_M}$, as seen in the second Section.

Let us hence take a different point of view: Leaving aside the local 
parametric Hamiltonian, we focus on the qubit state.
Once $M$ is fixed, the qubit is described by all the kets 
$\{\ket{\phi^\pm_M(\varphi)}\}_{\varphi\in[0,2\pi)}$, which define a 
mixed state. In fact, as these states are all equally likely, the 
corresponding density matrix is given by the ensemble 
average of their projectors, i.e. 
$\rho^{\pm;M}_\sigma=
\frac{1}{2\pi}\int\text{d}\varphi
\ket{\phi^\pm_M(\varphi)}\bra{\phi^\pm_M(\varphi)}$.
Due to the $\varphi$-integral, that makes the off-diagonal elements 
vanish, $\rho^{\pm;M}_\sigma$ coincides with the reduced 
density matrix of the qubit in the original fully quantum model, i.e.
${\text{Tr}}_{_{\rm ring}}\ket{\Psi^\pm_M}\bra{\Psi^\pm_M}$.
Therefore, not only the qubit can still be characterized by 
a finite Von Neumann entropy, $E^{\rm VN}(\rho_\sigma)$, but 
this turns out to be equal to the entanglement between 
the qubit and its environment before the quantum limit of the latter is 
taken, i.e.
\begin{equation}
{\cal{E}}_{\sigma S}(\ket{\Psi^\pm_M})
=
E^{\rm VN}(\rho^{\pm;M}_{\bm\sigma})~.
\label{e.entanglement-VonNeumann}
\end{equation}
The above statistical picture fits in the 
scheme proposed in Ref.\cite{HartleyV04} for evaluating the Von Neumann 
entropy of a qubit in whatever mixed state $\rho_\sigma$. It is there 
shown that when a qubit is described by an ensemble of $K$ equally 
likely pure states, corresponding to the Bloch vectors 
$\{{\bm n}_p\}_{p\subset P}$,  with $P$ some parameter space,
its density matrix is 
$\rho_\sigma=\frac{1}{2}(1+\overline{\bm n}\cdot{\bm\sigma})$, with the 
"average" Bloch vector $\overline{\bm n}$ defined as $\overline{\bm 
n}\equiv \frac{1}{K}\sum_p{\bm n}_p$, and its Von Neumann entropy can be 
written as
\begin{equation}
E^{\rm VN}(\rho_\sigma)=-h\left[\frac{1}{2}(1-|\overline{\bm n}|)\right]
\label{e.VonNeumann}
\end{equation}
Furthermore, given the ensemble $\{{\bm n}_p\}_{p\subset P}$, one can 
choose 
a sequence ${\bm n}_{p_1}\to{\bm n}_{p_2}\to{\bm n}_{p_3}...\to{\bm 
n}_{p_K}$ 
thus defining a curve in the parameter space $P$. To this curve
corresponds a Pancharatnam phase\cite{Pancharatnam56,Resta00} 
$\gamma\equiv{\rm arg}\{{\rm Tr}[\ket{{\bm n}_{p_1}}
\Pi_{i=1}^{K-1}\scalar{{\bm n}_{p_i}}{{\bm n}_{p_{i+1}}}
\bra{{\bm n}_{p_K}}]\}$, 
for which several relations with geometrical properties of the ensemble 
$\{{\bm n}_p\}_{p\subset P}$ can be found\cite{HartleyV04}.
Interpreting the sum defining $\overline{\bm n}$ as an ensemble 
average, and generalizing the above construction to the continuum case, 
we find that the mixed state in which the central qubit of the spin star is 
left, after the classical limit of the ring is taken, corresponds to 
\begin{equation}
\overline{\bm n}^\pm_M
\equiv\int d\varphi\,{\bm n}^\pm_M(\varphi)=(0,0,\pm\cos\vartheta_M)
\label{e.averagen}
\end{equation}
so that 
$E^{\rm VN}(\rho_\sigma^{\pm;M})=-h\left[\frac{1}{2}(1-\cos\vartheta_M)\right]$, 
consistently with Eqs.\eqref{e.entanglement-VonNeumann} and 
\eqref{e.entanglement}.
Furthermore, given our continuous ensemble, we can choose 
the closed sequence corresponding to the clock(-) or counterclock(+) wise 
oriented (with respect to the qubit quantization axis) 
line of latitude $\vartheta_M$, for which the Pancharatnam phase  
can be shown to be
\begin{equation}
\gamma_\pm=\pi(1\pm|\overline{\bm n}^\pm_M|)=\pi(1\pm\cos\vartheta_M)~,
\label{e.Gamma}
\end{equation}
which finally allows us to establish, via 
Eqs.\eqref{e.entanglement-VonNeumann} to \eqref{e.Gamma}, the relation
\begin{equation}
\label{e.entgamma}
\mathcal{E}_{\sigma S}=-h\left[\frac{\gamma_\pm}{2\pi}\right]\,\,.
\end{equation}
Notice that the phase $\gamma^\pm$ as obtained via the above 
construction is exactly the Berry phase which is picked by the 
eigenstates of a qubit in a magnetic field $\frac{g}{2}{\bm n}^\pm_M(\varphi)$ 
adiabatically rotating around the quantization axis with fixed polar 
angle $\vartheta_M$, a problem that pertains to the study of models with 
parametric Hamiltonians. However, in the usual approach to the study of these 
models the parametric dependence is 
not derived as the effect of some quantum environment, but rather assumed 
a priori, a viewpoint that leaves no space for entanglement properties,
hampering the disclosure of their possible relation with geometrical phases.

We underline that the relation Eq.\eqref{e.entgamma} specifically holds 
just for the particular model here considered. However, we believe that 
the reasoning behind its derivation has a more general content, that 
pertains to the analysis of the relation between geometrical properties 
of quantum systems and the structure, or dynamical evolution, of their 
states \cite{Levay04,Chruscinski06,AharonovA87,Sjoqvist00,TongEtal03, 
BertlmannEtal04,Basu06,Uhlmann86,SjoqvistEtal00,TongEtal04}. In 
particular we underline that the entanglement of the state of the 
original composite quantum system is the ultimate responsible for the 
dependence on the coeherent state variables to be conveyed from the 
environment to either the states or the local parametric Hamiltonian 
(depending on what point of view is adopted) of the principal system. 
In the limit of a classical environment, these variables become the 
parameters by exploring whose space the principal system can experience 
geometrical effects.

\section{Conclusions}
In this work we have proposed a method for studying the behaviour of an 
open quantum system along the quantum-to-classical crossover of its 
environment. The method, which is based on an exact, parametric 
representation for whatever state of an isolated system, is an original tool 
for dealing with phenomena which 
manifest themselves, and can be interpreted, very differently depending on the 
way the environment is modelled, not only in physical (see for 
instance Refs.\cite{PalmaSE96,LoFrancoEtal12,DarrigoEtal12}) but also in 
chemical and biological 
processes\cite{MohseniEtal08,PanitchayangkoonEtal2011,ScholesEtal11}.
As a first direct outcome, this work clarifies why
modelling a quantum system with a parametric Hamiltonian implies 
the existence of an environment ("the rest of the Universe" to use 
Berry's words\cite{Berry84}) and shows that a non-trivial 
parametric dependence can arise if and only if such environment is 
entangled with the system itself. 
One of the most relevant consequences of the above statement is that
the emergence of observable (i.e. gauge-invariant) quantities 
which are not eigenvalues of Hermitian operators of the system under 
analysis, such as the Berry's phase, turns out to be related not 
only to the fact that an environment exists \cite{Resta00}, 
but specifically to the condition that the system be entangled with its
environment. In fact, in the specific case of the spin-star here
considered, Eq.\eqref{e.entgamma} establishes that the entanglement
between the qubit and its environment can be determined by measuring the
observable Berry's phase characterizing the related model of a closed
qubit in a magnetic field, which suggests a possible way for
experimentally access entanglement properties via the observation of
gauge-invariant phases.

We conclude by mentioning some possible developments of this 
work that we think are worth pursuing. First of all, one might exploit 
the peculiar properties characterizing the dynamics of generalized coherent 
states, in order to relate a possible dynamical evolution of the global 
system to that of the principal one. Indeed, the 
formal scheme here presented 
opens the possibility of using established approaches for dealing 
with quantum dynamics in phase space, such as the the path-integral 
formalism, the adiabatic perturbation theory, the Born-Oppenheimer 
approximation, and generalizations to curved phase spaces of 
multi-configurational Eherefest methods,
as tools for taking into account the effects of the 
environment on the principal system and vice versa
\cite{KuratsujiI85,Kuratsuji88,Teufel03,LibertiEtal06,PanatiEtal07,YeSS12}.
Moreover, the approach here presented can be 
equivalently implemented in terms of generalized coherent states of any 
type\cite{ZhangFG90}, so that different physical systems can be taken into 
consideration, such as those belonging to the 
spin-bosons family.
As for the spin-$\frac{1}{2}$ star with frustration, it is worth 
mentioning that different types of interaction between 
the environmental spins, in particular the antiferromagnetic
Lieb-Mattis and Heisenberg-on-a-square-lattice ones, define exactly 
solvable models\cite{RichterEtal96} that can be treated in the very same 
framework here proposed. 
This expands the set of real physical systems where to look for 
a possible experimental analysis of our results.
Finally, we find particularly intriguing the idea of effectively studing 
the 
spin-$\frac{1}{2}$ star with frustration by quantum simulators
(see e.g. Ref.\cite{BulutaN09} and references therein,
and Refs.\cite{KimEtal10}):
In fact, the possibility of tuning the interaction parameters, 
which is recognized as one of the key features of quantum simulators, 
might allow the variation of the value of $S$,
by acting on the frustration ratio $k/g$, thus giving access to an 
experimental analysis of the crossover from a quantum to a classical 
environment.

\begin{acknowledgments}
The authors wish to thank T.J.G.Apollaro, L.Banchi, F.Bonechi, M.Long, 
A.Messina, M.Tarlini, and R.Vaia, for fruitful 
discussions and valuable suggestions. Financial support of the Italian
Ministry of Education, University, and Research in the framework
of the 2008 PRIN program (Contract No. 2008PARRTS
003) is also acknowledged.
\end{acknowledgments}


\end{document}